\documentclass[12pt]{iopart}
\usepackage{hyperr ef}
\usepackage{amssymb}
\hypersetup{colorlinks=true,linkcolor=blue,citecolor=blue,filecolor=blue,urlcolor=blue,breaklinks=true}
\RequirePackage{color}

\begin{document}
\title[Generalized relativistic harmonic oscillator in MLQM]{Generalized relativistic harmonic oscillator in minimal length quantum mechanics}

\author{L B Castro$^{1}$ and A E Obispo$^{1,2}$}

\address{$^1$ Departamento de F\'{\i}sica, Universidade Federal do Maranh\~{a}o (UFMA), Campus Universit\'{a}rio do Bacanga, 65080-805, S\~{a}o Lu\'{\i}s, MA, Brazil.}
\ead{luis.castro@pq.cnpq.br, lrb.castro@ufma.br}
\address{$^2$ Departamento de F\'{i}sica -- IGCE, Universidade Estadual Paulista (UNESP), Campus de Rio Claro, 13506-900, Rio Claro, SP, Brazil.}
\ead{signaux.fonce@gmail.com}

\begin{abstract}
We solve the generalized relativistic harmonic oscillator in $1+1$ dimensions in the presence of a minimal length. Using the momentum space representation, we explore all the possible signs of the potentials and discuss their bound-state solutions for fermions and antifermions. Furthermore, we also find an isolated solution from the Sturm--Liouville scheme. All cases already analyzed in the literature, are obtained as particular cases.
\end{abstract}

%
%
\submitto{\JPA}
%
%
%

\section{Introduction}
\label{intro}

The concept of a minimal measurable length scale, which is expected to be of
the order of the Planck length, given by $l_{p}=1.62\times 10^{-33}$cm, has
emerged as a condition required for a consistent formulation of quantum
theory of gravity \cite{LRR16:2:2013}. The existence of such minimal length,
which arise when quantum fluctuations of the gravitational field (at Planck
scale) are taken into account, is a common feature among most of theories
of quantum gravity such as, string theory \cite{PLB234:276:1990}, loop
quantum gravity \cite{IJMPA10:145:1995}, quantum cosmology \cite%
{PLB304:65:1993,JCAP2015:025:2015,NPB905:313:2016}, noncommutative field theories \cite{CTP61:605:2014,IJMPA29:1450106:2014,AP357:49:2015,AP362:24:2015}, and
black hole physics \cite%
{PLB452:39:1999,PRD81:023528:2010,IJMPA30:1550144:2015,EPL112:20006:2015,CQG33:025007:2016,IJTP55:617:2016}. One of
the interesting implications of introducing this minimal length is the
modification of the standard commutation relation between position and
momentum, which is transformed into a generalized relation that includes an
additional quadratic term in momentum, namely, $[X,P]=i\hbar \left( 1+\beta
P^{2}\right) $, where $\beta =\beta _{0}l_{p}/\hbar ^{2}$,$~\beta _{0}~$is a
dimensionless constant. This generalized commutation relation leads to the
modification of the Heisenberg uncertainty principle to a Generalized
Uncertainty Principle (GUP). This generalization, on the other hand, would
imply in the modification of the properties of the quantum system under
consideration, namely the eigenfunctions and the eigenvalues. This is the
main reason why recent works on the so-called Minimal Length Quantum
Mechanics (MLQM) has underwent a vertiginous growth \cite%
{PRD52:1108:1995,MPLA20:3095:2005,JPA39:5125:2006,PRL101:221301:2008,PS79:015010:2009, PLA373:1239:2009,CJP87:233:2009,PLA374:531:2010,PLB690:407:2010,PRD84:044013:2011, PLB718:678:2012, IJTP51:3963:2012,PLB711:423:2012,JPA45:385302:2012,ADHEP2013:383957:2013,PRD87:065017:2013,PLB718:1111:2013, EPJP128:25:2013,ADHEP2013:853696:2013,PRD89:105030:2014,PLB729:33:2014,AP355:269:2015, FBS56:19:2015,FP45:507:2015,IJMPA30:1550183:2015,PRD91:124017:2015,IJMPD25:1650013:2016,EPJC76:30:2016,
PLB755:17:2016,AP373:521:2016,PLB757:244:2016}%
.

The Dirac equation in 3+1 dimensions with a mixture of spherically symmetric scalar, vector and anomalous magnetic-like
(tensor) interactions can be reduced to the 1+1 Dirac equation with a mixture of scalar ($V_s$), time-component vector ($V_t$) and 
pseudoscalar ($V_p$) couplings when the fermion is limited to move in just one direction \cite{STRANGE1998}. In this restricted
 motion the scalar and vector interactions preserve their Lorentz structures while the anomalous magnetic-like interaction 
becomes a pseudoscalar. So, in the context of 1+1 dimensions the potential composed by $V_s$, $V_t$ and $V_p$ is the
most general combination of Lorentz structures because there are only four
linearly independent $2\times 2$ matrices. In a recent work published, Hassanabadi et al. \cite{PLB718:1111:2013} solved the minimal
length Dirac equation with harmonic oscillator potential (scalar and vector
interactions) and the energies and solutions are showed in a quite simple
and systematic manner. To the best of our knowledge, no one has reported on
the solution of the Dirac equation with a generalized relativistic harmonic
oscillator in $1+1$ dimensions in the presence of a minimal length and we
believe that this problem deserves to be explored.

In this context, the purpose of this work is to investigate the effects of
GUP when brought into problem of relativistic fermions moving in $1+1$
dimensions when a vector, scalar and pseudoscalar potentials are applied. We
consider the effect of GUP on definition of momentum operator and then we
obtain a generalized Dirac equation and solve exactly the corresponding
eigenvalue problem for the case of a generalized relativistic harmonic
oscillator. This problem is mapped into a Schr\"{o}dinger--like equation embedded in a 
symmetric P\"{o}schl-Teller potential. We explore all the possible signs of the potentials, thus paying attention to bound states of fermions and antifermions. Finally, we show that our results obtained for the energy
spectrum and the eigenfunctions are a generalization to those obtained in \cite{JPA39:5125:2006} (Dirac oscillator) and  \cite{PLB718:1111:2013} (mixed vector--scalar harmonic oscillator). Also, in the limit of the ordinary quantum mechanics ($\beta \rightarrow 0$) we are able to reproduce the case of generalized relativistic harmonic oscillator \cite{PRC73:054309:2006}.

\section{The generalized uncertainty principle}

We consider the following one-dimensional deformed commutation relation 
\begin{equation}\label{rgup}
\lbrack X,P]=i\hbar \left( 1+\beta P^{2}\right)\,,  
\end{equation}%
\noindent where $0\leqslant \beta \leqslant 1$. The limits $\beta
\rightarrow 0$ and $\beta \rightarrow 1$ correspond to the ordinary quantum
mechanics (OQM) and the extreme quantum gravity (EQG), respectively. This
deformed commutation relation leads to the following generalized uncertainty
principle (GUP) 
\begin{equation}\label{gup}
\Delta X\Delta P\geqslant \frac{\hbar }{2}\left[ 1+\beta (\Delta P)^{2}%
\right]\,.
\end{equation}%
\noindent The peculiarity of (\ref{gup}) is that it implies the existence of
a non-zero minimal uncertainty in position (minimal length). The
minimization of (\ref{gup}) with respect to $\Delta P$ gives 
\begin{equation}
\left( \Delta X\right) _{\mathrm{min}}=\hbar \sqrt{\beta }\,.  \label{ml}
\end{equation}%
\noindent Following \cite{PRD52:1108:1995}, we consider the following simple
one-dimensional realization of the position and momentum operators obeying
the relation (\ref{rgup}): 
\begin{equation}
X=i\hbar \left( 1+\beta p^{2}\right) \frac{\partial }{\partial p}\,,\qquad
P=p\,.  \label{repre}
\end{equation}%
\noindent It is important to note that the scalar product in this case is
not the usual one, but is defined as 
\begin{equation}
\langle f\mid g\rangle =\int_{-\infty }^{+\infty }\frac{dp}{(1+\beta p^{2})}%
f^{\ast }(p)g(p)\,.  \label{proi}
\end{equation}

\section{The Dirac equation in $1+1$ dimensions}

The $1+1$ dimensional time-independent Dirac equation for a fermion of rest
mass $m$ under the action of vector ($V_{t}$), scalar ($V_{s}$) and
pseudoscalar ($V_{p}$) potentials can be written, in terms of the
combinations $\Sigma =V_{t}+V_{s}$ and $\Delta =V_{t}-V_{s}$, as 
\begin{equation}\label{eq1}
H\psi =E\psi  \,,
\end{equation}
\noindent with 
\begin{equation}  \label{eq1b}
H=c\sigma _{1}P+\sigma _{3}mc^{2}+\frac{I+\sigma _{3}}{2}\Sigma +\frac{%
I-\sigma _{3}}{2}\Delta +\sigma _{2}V_{p}\,,
\end{equation}

\noindent where $E$ is the energy of the fermion and $P$ is the momentum
operator. The matrices $\sigma _{1}$, $\sigma_{2}$ and $\sigma_{3}$ denote
the Pauli matrices, and $I$ denotes the $2\times 2$ unit matrix. The
positive definite function $|\psi |^{2}=\psi ^{\dagger }\psi $, satisfying a
continuity equation, is interpreted as a position probability density and
its norm is a constant of motion. This interpretation is completely
satisfactory for single-particle states \cite{THALLER1992}.

\subsection{Equations of motion and isolated solution}

If we now write the spinor $\psi $ in terms of its components $\psi
^{T}=(f,\,g)$, the Dirac equation give rise to two coupled first-order
equations for the upper, $f$ and the lower, $g$ components of the spinor: 
\begin{equation}
\left[ cP-iV_{p}(X)\right]g=\left[E-mc^{2}-\Sigma(X)\right]f\,,
\label{eq11a}
\end{equation}%
\begin{equation}
\left[ cP+iV_{p}(X)\right]f=\left[E+mc^{2}-\Delta(X)\right]g\,.
\label{eq11b}
\end{equation}
\noindent In terms of $f$ and $g$ the spinor is normalized as 
\begin{equation}\label{condn1}
\int^{+\infty }_{-\infty }\frac{dP}{1+\beta P^{2}}(|f|^{2}+|g|^{2})=1\,,
\end{equation}
\noindent so that $f$ and $g$ are square integrable functions.

For $\Delta =0$ with $E\not=-mc^{2}$, the Dirac equation becomes 
\begin{equation}
g=\frac{(cP+iV_{p})f}{E+mc^{2}}\,,  \label{eq15a}
\end{equation}%
\begin{equation}
c^{2}P^{2}f-ic[V_{p},P]f+\left[ (E+mc^{2})\Sigma +V_{p}^{2}\right]
f=(E^{2}-m^{2}c^{4})f\,,  \label{eq15b}
\end{equation}%
\noindent and for $\Sigma =0$ with $E\not=mc^{2}$, the Dirac equation
becomes 
\begin{equation}
f=\frac{(cP-iV_{p})g}{E-mc^{2}}\,,  \label{eq16a}
\end{equation}%
\begin{equation}
c^{2}P^{2}g+ic[V_{p},P]g+\left[ (E-mc^{2})\Delta +V_{p}^{2}\right]
g=(E^{2}-m^{2}c^{4})g\,.  \label{eq16b}
\end{equation}%
\noindent Either for $\Delta =0$ with $E\not=-mc^{2}$ or $\Sigma =0$ with $%
E\not=mc^{2}$ the solution of the relativistic problem is mapped into a
Sturm-Liouville problem in such a way that solution can be found by solving
a Schr\"{o}dinger-like problem.

The solutions for $\Delta =0$ with $E=-mc^{2}$ and $\Sigma=0$ with $E=mc^{2}$
(isolated solution) can be obtained directly from the original first-order
equations (\ref{eq11a}) and (\ref{eq11b}). They are 
\begin{equation}  \label{isolada1a}
(cP-iV_{p}(X))g=\left[ -2mc^{2}-\Sigma(X) \right]f\,,
\end{equation}
\begin{equation}  \label{isolada1b}
(cP+iV_{p}(X))f=0\,,
\end{equation}
\noindent for $\Delta =0$ with $E=-mc^{2}$, and 
\begin{equation}  \label{isolada2a}
(cP-iV_{p}(X))g=0\,,
\end{equation}
\begin{equation}  \label{isolada2b}
(cP+iV_{p}(X))f=\left[ 2mc^{2}-\Delta(X) \right]g\,,
\end{equation}
\noindent for $\Sigma=0$ with $E=mc^{2}$. It is worthwhile to note that this
sort of isolated solution cannot describe scattering states and is subject
to the normalization condition (\ref{condn1}). Because $f$ and $g$ are
normalizable functions, the possible isolated solution presupposes $%
V_{p}(X)\neq0$.

\subsection{The nonrelativistic limit}

In the nonrelativistic limit (potential energies small compared to $mc^{2}$ and $E\sim mc^{2}$), the equation (\ref{eq15b}) becomes 
\begin{equation}\label{eqLNR}
\frac{P^{2}}{2m}f+\left( \frac{V_{p}^{2}}{2mc^{2}}-\frac{i\left[V_{p},P\right]}{2mc}+\Sigma \right)f=\mathcal{E}f\,,
\end{equation}
\noindent where $\mathcal{E}=E-mc^{2}$ is the nonrelativistic energy and $f$ obeys a Schr\"{o}dinger equation with a effective potential expressed in terms of the original potentials $\Sigma$ and $V_{p}$. Note that, in this approximation, $\Sigma$ preserves its original structures itself. However, $V_{p}$ provides two terms proportional to $i\left[V_{p},P\right]$ and $V_{p}^{2}$, which do not have a nonrelativistic counterpart. Therefore, we can say that $V_{p}$ is a potential intrinsically relativistic. Furthermore, the term $V_{p}^{2}$ in (\ref{eqLNR}) allows us to infer that even a potential unbounded from below could be a confining potential.

\section{The generalized relativistic harmonic oscillator}

Let us consider 
\begin{equation}\label{poten}
\Sigma =k_{1}X^{2},\qquad \Delta =0,\qquad V_{p}=k_{2}X\,. 
\end{equation}%
\noindent Note that, by making the changes $\Sigma \rightarrow \Delta $, $%
m\rightarrow -m$, $V_{p}\rightarrow -V_{p}$, $f\rightarrow g$ and $%
g\rightarrow f$ in Eq.(\ref{eq11a}) [or Eq. (\ref{eq15b})], we obtain Eq. (%
\ref{eq11b}) [or Eq. (\ref{eq16b})]. This symmetry also is present in the
isolated solution and can be clearly seen from the two equation pairs (\ref%
{isolada1a})-(\ref{isolada1b}) and (\ref{isolada2a})-(\ref{isolada2b}). This
symmetry provides a simple mechanism by which one can go from the results
from the case $\Sigma =k_{1}X^{2}$, $\Delta =0$, $V_{p}=k_{2}X$ to the case
when $\Delta =k_{1}X^{2}$, $\Sigma =0$, $V_{p}=k_{2}X$ by just changing the
sign of $m$ and of $k_{2}$ in the relevant expressions.

The configuration for the potentials (\ref{poten}) was chosen conveniently so that one obtains Schr\"{o}dinger-like equations with the harmonic
 oscillator potential. We can note that, for the case $\Delta=0$ ($V_{t}=V_{s}$) and a confining vector potential ($\sim X^{2}$), the scalar potential $V_{s}$ is also a confining potential. In this case, the $\Sigma$ potential, which appears in the Schr\"{o}dinger-like equation for the component $f$ [Eq. (\ref{eq15b})] becomes proportional to $X^{2}$ (confining potential). For $\Sigma=0$ ($V_{t}=-V_{s}$) and a confining vector potential ($\sim X^{2}$), the scalar potential is not a confining potential ($\sim-X^{2}$). In this case, the $\Delta$ potential, which appears in the Schr\"{o}dinger-like equation for the component $g$ [Eq. (\ref{eq16b})] also becomes proportional to $X^{2}$ (confining potential). In both cases for a linear pseudoscalar potential, we always have a confining potential in the Schr\"{o}dinger-like equation for $f$ and $g$ components. The linear pseudoscalar potential is associated to a well known system, the Dirac oscillator \cite{JPA22:L817:1989}. The Dirac oscillator is a natural model for studying properties of physical systems, it is an exactly solvable model, several research have been developed in the context of this theoretical framework in recent years \cite{STRANGE1998}.

\subsection{Isolated solution}

The isolated solution with $E=-mc^{2}$ is obtained from Eqs. (\ref{isolada1a}%
) and (\ref{isolada1b}). Substituting (\ref{poten}) in Eqs. (\ref{isolada1a}%
) and (\ref{isolada1b}) and using the operator relation (\ref{repre}), we
obtain 
\begin{equation}  \label{eq20a}
\frac{dg(p)}{dp}+\frac{cp}{\hbar k_{2}\left(1+\beta p^{2}\right)}%
g(p)=Q(p)f(p)\,,
\end{equation}
\noindent and 
\begin{equation}  \label{eq20b}
\frac{df(p)}{dp}-\frac{cp}{\hbar k_{2}\left(1+\beta p^{2}\right)}f(p)=0\,,
\end{equation}
\noindent respectively. It is important to mention that the expression for $%
Q(p)$ is irrelevant to our analysis. The general solutions for (\ref{eq20a})
and (\ref{eq20b}) are given by 
\begin{eqnarray}
f(p) &=& N_{+}\left( 1+\beta p^{2} \right)^{\frac{c}{2\beta \hbar k_{2}}},
\label{isoladaf} \\
g(p) &=& \left( 1+\beta p^{2} \right)^{-\frac{c}{2\beta \hbar k_{2}}}\left[
N_{+}I(p)+N_{-} \right],  \label{isoladag}
\end{eqnarray}%
\noindent where $N_{+}$ and $N_{-}$ are normalization constants, and 
\begin{equation}  \label{defI}
I(p)=\int{Q(p)\left( 1+\beta p^{2} \right)^{\frac{c}{\beta \hbar k_{2}}}dp}%
\,.
\end{equation}
\noindent Observing (\ref{isoladaf}) and (\ref{isoladag}), we can conclude
that it is impossible to have both nonzero components simultaneously as
physically acceptable solution. A normalizable solution for $k_{2}>0$ and $%
\beta\neq 0$ is possible if $N_{+}=0$. Thus, 
\begin{equation}  \label{sol1}
\psi(p)=N(1+\beta p^{2})^{-\frac{c}{2\beta \hbar k_{2}}} \left( 
\begin{array}{c}
0 \\ 
1 \\ 
\end{array}
\right)\,,
\end{equation}
\noindent where 
\begin{equation}  \label{consnorma}
N=\sqrt{\frac{1}{\delta}}\,,
\end{equation}
\noindent with 
\begin{equation}  \label{nconst}
\delta=\int^{+\infty}_{-\infty}{\frac{dp}{(1+\beta p^{2})^{\left(\frac{\beta
\hbar k_{2}+c}{\beta \hbar k_{2}}\right)}}}\,.
\end{equation}
\noindent

\subsection{Solution for $\Delta=0$ and $E\neq-mc^{2}$}

For $\Delta =0$ and $E\neq -mc^{2}$, the Eq. (\ref{eq15b}) takes the form 
\begin{equation}\label{eq2of}
\left( c^{2}+\hbar c\beta k_{2}\right) P^{2}f(P)+\kappa
^{2}X^{2}f(P)=\varepsilon ^{2}f(P)\,, 
\end{equation}%
\noindent where 
\begin{equation}
\kappa ^{2}=\left( E+mc^{2}\right) k_{1}+k_{2}^{2}\,,  \label{kappa}
\end{equation}%
\begin{equation}
\varepsilon ^{2}=E^{2}-m^{2}c^{4}-\hbar ck_{2}\,,  \label{eeff}
\end{equation}%
\noindent and the Eq. (\ref{eq15a}) becomes
\begin{equation}\label{solgpo}
g(P)=\frac{cP+ik_{2}X}{E+mc^{2}}f(P)\,.
\end{equation}
\noindent Using the operator relation (\ref{repre}), the Eq. (\ref{eq2of})
gets 
\begin{equation}
\frac{d^{2}f(p)}{dp^{2}}+\frac{2\beta p}{1+\beta p^{2}}\frac{df(p)}{dp} 
 +\frac{\eta ^{2}}{(1+\beta p^{2})^{2}}\left[ \varepsilon ^{2}-(c^{2}+\hbar
c\beta k_{2})p^{2}\right] f(p)=0\,,
\label{eq21}
\end{equation}%
\noindent where $\eta ^{2}=\frac{1}{\hbar ^{2}\kappa ^{2}}$, and (\ref{solgpo}) becomes
\begin{equation}\label{solgpo2}
g(p)=-\frac{\hbar k_{2}\left(1+\beta p^{2}\right)}{E+mc^{2}}\left[ \frac{d}{dp}-\frac{cp}{\hbar k_{2}\left(1+\beta p^{2}\right)} \right]f(p)\,.
\end{equation}
\noindent Implementing a
change of variable defined by 
\begin{equation}\label{vacha}
q=\frac{1}{\hbar \sqrt{\beta }\kappa }\arctan{\left(\sqrt{\beta }p\right)}\,,
\end{equation}%
\noindent we can rewrite the Eq. (\ref{eq21}) as 
\begin{equation}
\frac{d^{2}f(q)}{dq^{2}}-\frac{c^{2}+\hbar c\beta k_{2}}{\beta }\tan ^{2}{%
\left(\hbar \sqrt{\beta }\kappa q\right)}f(q)+\varepsilon ^{2}f(q)=0.  \label{eq22}
\end{equation}%
\noindent where we recognize the effective potential as the exactly solvable
symmetric P\"{o}schl-Teller potential \cite{JPA40:263:2007} with $k_{2}>-\frac{c}{\hbar\beta}$. The
corresponding effective eigenenergy is given by 
\begin{equation}\label{energiaeff}
\frac{\varepsilon ^{2}}{\hbar ^{2}\beta \kappa ^{2}}=n\left( n+2\lambda
\right) +\lambda\,,   
\end{equation}%
\noindent where $n$ is a non-negative integer and 
\begin{equation}
\lambda =\frac{1}{2}+\frac{1}{2\hbar \beta }\sqrt{\frac{\hbar ^{2}\beta
^{2}(E+mc^{2})k_{1}+(\hbar \beta k_{2}+2c)^{2}}{(E+mc^{2})k_{1}+k_{2}^{2}}}%
\,.  \label{lambda}
\end{equation}%
\noindent Now, (\ref{eeff}),(\ref{energiaeff}) and (\ref{lambda}) tell us
that 
\begin{eqnarray}
\fl E^{2}=\hbar ^{2}\beta \left[ (E+mc^{2})k_{1}+k_{2}^{2}\right] \left(
n^{2}+n+\frac{1}{2}\right) \nonumber\\
+\left( n+\frac{1}{2}\right) \hbar \frac{\sqrt{\hbar ^{2}\beta
^{2}(E+mc^{2})k_{1}+(\hbar \beta k_{2}+2c)^{2}}}{\left[
(E+mc^{2})k_{1}+k_{2}^{2}\right] ^{-1/2}}
+m^{2}c^{4}+\hbar ck_{2}\,.
\label{spectrum}
\end{eqnarray}
\noindent The solutions of (\ref{spectrum}) determine the eigenvalues of our
problem. This equation can be solved easily with a symbolic algebra program.

The solution for the differential equation (\ref{eq22}) becomes
\begin{equation}\label{solgegen}
f(q)=A\left(\cos\left(\hbar\sqrt{\beta}\kappa q\right)\right)^{\lambda}
C_{n}^{\lambda}\left(\sin\left(\hbar\sqrt{\beta}\kappa q\right)\right)\,,
\end{equation}
\noindent where $A$ is a normalization constant and $C_{n}^{\lambda}(z)$ are the Gegenbauer polynomials \cite{ABRAMOWITZ1965}. The lower component obtained from (\ref{solgpo2}) is given by
\begin{equation}\label{solg}
g(q)=-\frac{2iAc\left(\cos\left(\hbar\sqrt{\beta}\kappa q\right)\right)^{\lambda+1}}{\sqrt{\beta}\left(E+mc^2\right)}C_{n-1}^{\lambda+1}\left(\sin\left(\hbar\sqrt{\beta}\kappa q\right)\right)\,.
\end{equation}
\noindent Using the relation (\ref{vacha}) the solutions (\ref{solgegen}) and (\ref{solg}) can be rewritten in function of the original variable $p$ as
\begin{equation}\label{sol1p}
f(p)=A\left(1+\beta p^{2}\right)^{-\lambda/2}C_{n}^{\lambda}
\left( \frac{p\sqrt{\beta}}{\sqrt{1+\beta p^{2}}} \right)\,,
\end{equation}
\noindent and
\begin{equation}\label{sol2p}
g(p)=-\frac{2iAc\left(1+\beta p^{2}\right)^{-\lambda-1}}{\sqrt{\beta}\left(E+mc^{2}\right)}
C_{n-1}^{\lambda+1}\left( \frac{p\sqrt{\beta}}{\sqrt{1+\beta p^{2}}} \right)\,,
\end{equation}
\noindent respectively. The normalization condition (\ref{condn1}) dictates that the normalization constant can be written as
\begin{equation}\label{constnorm}
\fl A=\frac{2^{\lambda}\beta^{1/4}}{\sqrt{2\pi}}\left[ \frac{\Gamma(2\lambda+n)}{n!(n+\lambda)\left(\Gamma(\lambda)\right)^{2}}
+\frac{c^{2}}{\beta\left(E+mc^{2}\right)^{2}}\frac{\Gamma(2\lambda+n+1)}{(n-1)!(n+\lambda)\left(\Gamma(\lambda+1)\right)^{2}}\right]^{-1/2}\,.
\end{equation}

\section{Particular cases on minimal length quantum mechanics (MLQM)}

\subsection{Pure pseudoscalar linear potential (one-dimensional Dirac
oscillator)}

For $\beta \neq 0$ and $k_{1}=0$ the expression (\ref{spectrum}) reduces to 
\begin{equation}\label{case4}
E^{2}-m^{2}c^{4} =\hbar ^{2}\beta k_{2}^{2}\left( n^{2}+n+\frac{1}{2} \right)
+\hbar\left( n+\frac{1}{2} \right)|\hbar\beta k_{2}+2c||k_{2}|+\hbar ck_{2} \,.
\end{equation}%
\noindent One can readily envisage that two different classes of solutions can be distinguished depending on the sign of $k_{2}$. 

For $k_{2}>0$, we obtain 
\begin{equation}\label{enerk2+}
E=\pm \sqrt{m^{2}c^{4}+2\left( n+1\right) \hbar ck_{2}+\beta \hbar
^{2}k_{2}^{2}\left( n+1\right) ^{2}}\,.  
\end{equation}%
\noindent For $-\frac{c}{\hbar\beta}<k_{2}<0$, we obtain 
\begin{equation}\label{enerk2-}
E=\pm \sqrt{m^{2}c^{4}+2n\hbar c|k_{2}|+\beta \hbar ^{2}k_{2}^{2}n^{2}}\,.
\end{equation}%
\noindent This last result is exactly the Eq. (37) of Ref. \cite%
{JPA39:5125:2006}. Note that $n\geqslant0$ for $k_{2}>0$ and $n\geqslant1$ for $-\frac{c}{\hbar\beta}<k_{2}<0$, because for the lower component is proportional to a Gegenbauer polynomial of degree $n-1$. Our results (\ref{enerk2+}) and (\ref{enerk2-}), allow us
to conclude that 
\begin{equation}\label{enerk2}
E_{\pm }=\pm \sqrt{m^{2}c^{4}+2\left( n+1\right) \hbar c|k_{2}|+\beta \hbar
^{2}k_{2}^{2}\left( n+1\right) ^{2}}\,,  
\end{equation}%
\noindent with $n=0,1,\ldots $, and it is independent of the sign of $k_{2}$%
. Note that both particle ($E_{+}$) and antiparticle ($E_{-}$) energy levels
are members of the spectrum.

Let us consider the limit of the ordinary quantum mechanics ($\beta
\rightarrow 0$). In this limit (\ref{enerk2}) becomes 
\begin{equation}
E_{\pm }=\pm \sqrt{m^{2}c^{4}+2\left( n+1\right) \hbar c|k_{2}|}\,,
\label{enerk20}
\end{equation}%
\noindent which is in accordance with Ref. \cite{PRC73:054309:2006}.

\subsection{Mixed vector--scalar harmonic oscillator potentials}

For $\beta \neq 0$ and $k_{2}=0$ the expression (\ref{spectrum}) reduces to 
\begin{equation}
\fl E^{2}-m^{2} =\left( n+\frac{1}{2}\right) \hbar \frac{\sqrt{\hbar ^{2}\beta
^{2}(E+mc^{2})k_{1}+4c^{2}}}{\left[ (E+mc^{2})k_{1}\right] ^{-1/2}}
+\hbar ^{2}\beta \left( n^{2}+n+\frac{1}{2}\right) \left[ (E+mc^{2})k_{1}%
\right] \,.
\label{case5}
\end{equation}%
\noindent This result is equivalent to Eq. (12) of Ref. \cite%
{PLB718:1111:2013}. The quantization condition expressed by (\ref{case5})
can be rewritten as 
\begin{eqnarray}
\fl (E-mc^{2})\sqrt{\left( E+mc^{2}\right) \mathrm{sgn}(k_{1})}= \nonumber\\
 \mathrm{sgn}(k_{1})\left[ \frac{\left( n+\frac{1}{2}\right) \hbar \sqrt{%
|k_{1}|}}{\left[ \hbar ^{2}\beta ^{2}(E+mc^{2})\mathrm{sgn}(k_{1})|k_{1}|+4c^{2}%
\right] ^{-1/2}}\right.  \nonumber\\
 \left. +\hbar ^{2}\beta \left( n^{2}+n+\frac{1}{2}\right) \sqrt{\left(
E+mc^{2}\right) \mathrm{sgn}(k_{1})}|k_{1}|\right] \,.
\label{enerk1}
\end{eqnarray}%
\noindent This result implies that when $k_{1}>0$ there are only bound states for
fermions with $E>mc^{2}$. On the other hand, for $k_{1}<0$ there are only
bound states for antifermions with $E<-mc^{2}$. Therefore, the positive
(negative) energies for fermions (antifermions) never sink into Dirac sea of
negative (positive) energies. This fact means that there is no channel for
spontaneous fermion--antifermion creation (Klein's paradox).

Now, let us consider the limit of the ordinary quantum mechanics ($\beta
\rightarrow 0$). In this limit (\ref{enerk1}) becomes 
\begin{equation}
(E-mc^{2})\sqrt{\left( E+mc^{2}\right) \mathrm{sgn}(k_{1})}=\mathrm{sgn}%
(k_{1})\left( n+\frac{1}{2}\right) \hbar c\sqrt{4|k_{1}|}\,,  \label{enerk10}
\end{equation}%
\noindent which is in accordance with Ref. \cite{PRC73:054309:2006}.

\subsection{Generalized relativistic harmonic oscillator (ordinary quantum mechanics)}

Now, let us consider the limit of the ordinary quantum mechanics ($\beta\rightarrow0$). In this limit 
(\ref{spectrum}) reduces 
\begin{equation}\label{energho}
E^{2}-m^{2}c^{4}=\left(2n+1\right)\hbar c\sqrt{\left(E+mc^{2}\right)k_{1}+k_{2}^{2}}
+\hbar ck_{2}\,.
\end{equation}
\noindent This result is exactly the Eq. (44) of Ref. \cite{PRC73:054309:2006}.

\section{Final remarks}

In this paper, we have studied the problem of relativistic fermions moving
in $1+1$ dimensions under the influence of mixture of a vector, scalar and
pseudoscalar potentials (most general Lorentz structure) in the context of a Minimal Length Quantum
Mechanics. Using the momentum space representation and a convenient
representation of the Dirac matrices, we solved the first order Dirac
equation and found a isolated solution for $\Delta =0$ with $E=-mc^{2}$ for
the case of a generalized relativistic harmonic oscillator. It is important
to highlight that those solutions only exist for $\beta \neq 0$ and are
normalizable when $k_{2}>0$. Normalizable solutions for $\Sigma =0$ with $%
E=mc^{2}$ can be easily obtained from the symmetries of the system, as
mentioned above.

The expression for the energy spectrum and the corresponding eigenstates for 
$\Delta=0$ and $E\neq -mc^{2}$ were obtained exactly from the second--order differential equation for
the Dirac spinor components after being mapped into a Sturm-Liouville problem
(Schr\"{o}dinger--like) with a symmetric P\"{o}schl--Teller potential. For the general case, the quantization condition can be solved easily with a symbolic algebra program and the Dirac spinor were expressed in terms of the Gegenbauer polynomials. We discussed in detail all the possible signs of the potentials and determined which values of $k_{1}$ and $k_{2}$ allow there to be a spectrum of both fermion and antifermion bounded solutions simultaneously or just one of these type of solutions. Furthermore, a remarkable feature of this problem is that we were able to reproduce well--known particular cases of relativistic harmonic oscillator in the presence of a minimal length, as for instance: the cases of harmonic oscillator potential (scalar and vector couplings) \cite{PLB718:1111:2013} and the so--called Dirac oscillator \cite{JPA39:5125:2006}. Also, the results obtained in this work are consistent in the limit of the ordinary quantum mechanics ($\beta\rightarrow0$) with those found in \cite{PRC73:054309:2006} for the generalized relativistic harmonic oscillator. 

\ack
This work was supported in part by means of funds provided by CNPq (grants
455719/2014-4 and 304105/2014-7). Angel E. Obispo thanks to CAPES for support through a scholarship under the CAPES/PNPD program. Angel E. Obispo also thanks to CNPq (grant 312838/2016-6) and Secti/FAPEMA (grant FAPEMA DCR-02853/16), for financial support.

\section*{References}
\bibliography{mybibfile_stars2}

\providecommand{\newblock}{}
\begin{thebibliography}{10}
\expandafter\ifx\csname url\endcsname\relax
  \def\url#1{{\tt #1}}\fi
\expandafter\ifx\csname urlprefix\endcsname\relax\def\urlprefix{URL }\fi
\providecommand{\eprint}[2][]{\url{#2}}

\bibitem{LRR16:2:2013}
Hossenfelder S 2013 {\em Living Rev. Relativity\/} {\bf 16}
  \urlprefix\url{http://www.livingreviews.org/lrr-2013-2}

\bibitem{PLB234:276:1990}
Konishi K, Paffuti G and Provero P 1990 {\em Phys. Lett. B\/} {\bf 234} 276
  ISSN 0370-2693
  \urlprefix\url{http://www.sciencedirect.com/science/article/pii/0370269390919274}

\bibitem{IJMPA10:145:1995}
GARAY L~J 1995 {\em Int. J. Mod. Phys. A\/} {\bf 10} 145 (\textit{Preprint}
  \eprint{http://www.worldscientific.com/doi/pdf/10.1142/S0217751X95000085})
  \urlprefix\url{http://www.worldscientific.com/doi/abs/10.1142/S0217751X95000085}

\bibitem{PLB304:65:1993}
Maggiore M 1993 {\em Phys. Lett. B\/} {\bf 304} 65 ISSN 0370-2693
  \urlprefix\url{http://www.sciencedirect.com/science/article/pii/0370269393914018}

\bibitem{JCAP2015:025:2015}
Ali A~F, Faizal M and Khalil M~M 2015 {\em JCAP\/} {\bf 2015} 025
  \urlprefix\url{http://stacks.iop.org/1475-7516/2015/i=09/a=025}

\bibitem{NPB905:313:2016}
Garattini R and Faizal M 2016 {\em Nucl. Phys. B\/} {\bf 905} 313 ISSN
  0550-3213
  \urlprefix\url{http://www.sciencedirect.com/science/article/pii/S0550321316000675}

\bibitem{CTP61:605:2014}
Bing-Sheng L, Tai-Hua H and Wei C 2014 {\em Commun. Theor. Phys.\/} {\bf 61}
  605 \urlprefix\url{http://stacks.iop.org/0253-6102/61/i=5/a=11}

\bibitem{IJMPA29:1450106:2014}
Faizal M 2014 {\em Int. J. Mod. Phys. A\/} {\bf 29} 1450106 (\textit{Preprint}
  \eprint{http://www.worldscientific.com/doi/pdf/10.1142/S0217751X14501061})
  \urlprefix\url{http://www.worldscientific.com/doi/abs/10.1142/S0217751X14501061}

\bibitem{AP357:49:2015}
Faizal M and Majumder B 2015 {\em Ann. Phys. (N.Y.)\/} {\bf 357} 49 ISSN
  0003-4916
  \urlprefix\url{http://www.sciencedirect.com/science/article/pii/S0003491615001232}

\bibitem{AP362:24:2015}
Pramanik S, Moussa M, Faizal M and Ali A~F 2015 {\em Ann. Phys. (N.Y.)\/} {\bf
  362} 24 ISSN 0003-4916
  \urlprefix\url{http://www.sciencedirect.com/science/article/pii/S0003491615002912}

\bibitem{PLB452:39:1999}
Scardigli F 1999 {\em Phys. Lett. B\/} {\bf 452} 39 ISSN 0370-2693
  \urlprefix\url{http://www.sciencedirect.com/science/article/pii/S0370269399001677}

\bibitem{PRD81:023528:2010}
Bina A, Jalalzadeh S and Moslehi A 2010 {\em Phys. Rev. D\/} {\bf 81}(2) 023528
  \urlprefix\url{http://link.aps.org/doi/10.1103/PhysRevD.81.023528}

\bibitem{IJMPA30:1550144:2015}
Faizal M and Khalil M~M 2015 {\em Int. J. Mod. Phys. A\/} {\bf 30} 1550144
  (\textit{Preprint}
  \eprint{http://www.worldscientific.com/doi/pdf/10.1142/S0217751X15501444})
  \urlprefix\url{http://www.worldscientific.com/doi/abs/10.1142/S0217751X15501444}

\bibitem{EPL112:20006:2015}
Gangopadhyay S, Dutta A and Faizal M 2015 {\em EPL\/} {\bf 112} 20006
  \urlprefix\url{http://stacks.iop.org/0295-5075/112/i=2/a=20006}

\bibitem{CQG33:025007:2016}
Wang P, Yang H and Ying S 2016 {\em Class. Quant. Grav.\/} {\bf 33} 025007
  \urlprefix\url{http://stacks.iop.org/0264-9381/33/i=2/a=025007}

\bibitem{IJTP55:617:2016}
Gangopadhyay S 2016 {\em Int. J. Theor. Phys.\/} {\bf 55} 617 ISSN 1572-9575
  \urlprefix\url{http://dx.doi.org/10.1007/s10773-015-2699-7}

\bibitem{PRD52:1108:1995}
Kempf A, Mangano G and Mann R~B 1995 {\em Phys. Rev. D\/} {\bf 52}(2) 1108
  \urlprefix\url{http://link.aps.org/doi/10.1103/PhysRevD.52.1108}

\bibitem{MPLA20:3095:2005}
NOZARI K and KARAMI M 2005 {\em Mod. Phys. Lett. A\/} {\bf 20} 3095
  (\textit{Preprint}
  \eprint{http://www.worldscientific.com/doi/pdf/10.1142/S0217732305018517})
  \urlprefix\url{http://www.worldscientific.com/doi/abs/10.1142/S0217732305018517}

\bibitem{JPA39:5125:2006}
Nouicer K 2006 {\em J. Phys. A: Math. and Gen.\/} {\bf 39} 5125
  \urlprefix\url{http://stacks.iop.org/0305-4470/39/i=18/a=025}

\bibitem{PRL101:221301:2008}
Das S and Vagenas E~C 2008 {\em Phys. Rev. Lett.\/} {\bf 101}(22) 221301
  \urlprefix\url{http://link.aps.org/doi/10.1103/PhysRevLett.101.221301}

\bibitem{PS79:015010:2009}
Merad M and Falek M 2009 {\em Phys. Scr.\/} {\bf 79} 015010
  \urlprefix\url{http://stacks.iop.org/1402-4896/79/i=1/a=015010}

\bibitem{PLA373:1239:2009}
Jana T and Roy P 2009 {\em Phys. Lett. A\/} {\bf 373} 1239 ISSN 0375-9601
  \urlprefix\url{http://www.sciencedirect.com/science/article/pii/S0375960109001716}

\bibitem{CJP87:233:2009}
Das S and Vagenas E~C 2009 {\em Can. J. Phys.\/} {\bf 87} 233
  (\textit{Preprint} \eprint{http://dx.doi.org/10.1139/P08-105})
  \urlprefix\url{http://dx.doi.org/10.1139/P08-105}

\bibitem{PLA374:531:2010}
Chargui Y, Trabelsi A and Chetouani L 2010 {\em Phys. Lett. A\/} {\bf 374} 531
  ISSN 0375-9601
  \urlprefix\url{http://www.sciencedirect.com/science/article/pii/S0375960109014601}

\bibitem{PLB690:407:2010}
Das S, Vagenas E~C and Ali A~F 2010 {\em Phys. Lett. B\/} {\bf 690} 407 ISSN
  0370-2693
  \urlprefix\url{http://www.sciencedirect.com/science/article/pii/S0370269310006441}

\bibitem{PRD84:044013:2011}
Ali A~F, Das S and Vagenas E~C 2011 {\em Phys. Rev. D\/} {\bf 84}(4) 044013
  \urlprefix\url{http://link.aps.org/doi/10.1103/PhysRevD.84.044013}

\bibitem{PLB718:678:2012}
Hassanabadi H, Zarrinkamar S and Maghsoodi E 2012 {\em Phys. Lett. B\/} {\bf
  718} 678 ISSN 0370-2693
  \urlprefix\url{http://www.sciencedirect.com/science/article/pii/S0370269312011562}

\bibitem{IJTP51:3963:2012}
Ta{\c{s}}k{\i}n F and Yaman Z 2012 {\em Int. J. Theor. Phys.\/} {\bf 51} 3963
  ISSN 1572-9575 \urlprefix\url{http://dx.doi.org/10.1007/s10773-012-1288-2}

\bibitem{PLB711:423:2012}
Ghosh S and Roy P 2012 {\em Phys. Lett. B\/} {\bf 711} 423 ISSN 0370-2693
  \urlprefix\url{http://www.sciencedirect.com/science/article/pii/S0370269312004364}

\bibitem{JPA45:385302:2012}
Dey S, Fring A and Gouba L 2012 {\em J. Phys. A: Math. and Theor.\/} {\bf 45}
  385302 \urlprefix\url{http://stacks.iop.org/1751-8121/45/i=38/a=385302}

\bibitem{ADHEP2013:383957:2013}
Betrouche M, Maamache M and Choi J~R 2013 {\em AdHEP\/} {\bf 2013} Article ID
  383957 ISSN 1687-7357
  \urlprefix\url{http://www.hindawi.com/journals/ahep/2013/383957/cta/}

\bibitem{PRD87:065017:2013}
Menculini L, Panella O and Roy P 2013 {\em Phys. Rev. D\/} {\bf 87}(6) 065017
  \urlprefix\url{http://link.aps.org/doi/10.1103/PhysRevD.87.065017}

\bibitem{PLB718:1111:2013}
Hassanabadi H, Zarrinkamar S and Rajabi A 2013 {\em Phys. Lett. B\/} {\bf 718}
  1111 ISSN 0370-2693
  \urlprefix\url{http://www.sciencedirect.com/science/article/pii/S0370269312012105}

\bibitem{EPJP128:25:2013}
Hassanabadi H, Zarrinkamar S and Maghsoodi E 2013 {\em Eur. Phys. J. Plus\/}
  {\bf 128} 25 ISSN 2190-5444
  \urlprefix\url{http://dx.doi.org/10.1140/epjp/i2013-13025-1}

\bibitem{ADHEP2013:853696:2013}
Pedram P 2013 {\em AdHEP\/} {\bf 2013} Article ID 853696 ISSN 1687-7357
  \urlprefix\url{http://www.hindawi.com/journals/ahep/2013/853696/}

\bibitem{PRD89:105030:2014}
Haouat S and Nouicer K 2014 {\em Phys. Rev. D\/} {\bf 89}(10) 105030
  \urlprefix\url{http://link.aps.org/doi/10.1103/PhysRevD.89.105030}

\bibitem{PLB729:33:2014}
Haouat S 2014 {\em Phys. Lett. B\/} {\bf 729} 33 ISSN 0370-2693
  \urlprefix\url{http://www.sciencedirect.com/science/article/pii/S0370269313010356}

\bibitem{AP355:269:2015}
Bouaziz D 2015 {\em Ann. Phys. (N.Y)\/} {\bf 355} 269 ISSN 0003-4916
  \urlprefix\url{http://www.sciencedirect.com/science/article/pii/S0003491615000354}

\bibitem{FBS56:19:2015}
Hassanabadi H, Hooshmand P and Zarrinkamar S 2015 {\em Few-Body Syst.\/} {\bf
  56} 19 ISSN 1432-5411
  \urlprefix\url{http://dx.doi.org/10.1007/s00601-014-0910-7}

\bibitem{FP45:507:2015}
Falek M, Merad M and Moumni M 2015 {\em Found. Phys.\/} {\bf 45} 507 ISSN
  1572-9516 \urlprefix\url{http://dx.doi.org/10.1007/s10701-015-9880-y}

\bibitem{IJMPA30:1550183:2015}
Faizal M, Ali A~F and Nassar A 2015 {\em Int. J. Mod. Phys. A\/} {\bf 30}
  1550183 (\textit{Preprint}
  \eprint{http://www.worldscientific.com/doi/pdf/10.1142/S0217751X15501833})
  \urlprefix\url{http://www.worldscientific.com/doi/abs/10.1142/S0217751X15501833}

\bibitem{PRD91:124017:2015}
Dey S and Hussin V 2015 {\em Phys. Rev. D\/} {\bf 91}(12) 124017
  \urlprefix\url{https://link.aps.org/doi/10.1103/PhysRevD.91.124017}

\bibitem{IJMPD25:1650013:2016}
Faizal M and Kruglov S~I 2016 {\em Int. J. Mod. Phys. D\/} {\bf 25} 1650013
  (\textit{Preprint}
  \eprint{http://www.worldscientific.com/doi/pdf/10.1142/S0218271816500139})
  \urlprefix\url{http://www.worldscientific.com/doi/abs/10.1142/S0218271816500139}

\bibitem{EPJC76:30:2016}
Faizal M, Khalil M~M and Das S 2016 {\em Eur. Phys. J. C\/} {\bf 76} 30 ISSN
  1434-6052 \urlprefix\url{http://dx.doi.org/10.1140/epjc/s10052-016-3884-4}

\bibitem{PLB755:17:2016}
Deb S, Das S and Vagenas E~C 2016 {\em Phys. Lett. B\/} {\bf 755} 17 ISSN
  0370-2693
  \urlprefix\url{http://www.sciencedirect.com/science/article/pii/S0370269316000782}

\bibitem{AP373:521:2016}
Bernardo R~C~S and Esguerra J~P~H 2016 {\em Ann. Phys. (N.Y.)\/} {\bf 373} 521
  ISSN 0003-4916
  \urlprefix\url{http://www.sciencedirect.com/science/article/pii/S0003491616301427}

\bibitem{PLB757:244:2016}
Faizal M 2016 {\em Phys. Lett. B\/} {\bf 757} 244 ISSN 0370-2693
  \urlprefix\url{http://www.sciencedirect.com/science/article/pii/S0370269316300454}

\bibitem{STRANGE1998}
Strange P 1998 {\em Relativistic Quantum Mechanics with Applications in
  Condensed Matter and Atomic Physics\/} (Cambridge: Cambridge University
  Press) \urlprefix\url{http://dx.doi.org/10.1017/CBO9780511622755}

\bibitem{PRC73:054309:2006}
de~Castro A~S, Alberto P, Lisboa R and Malheiro M 2006 {\em Phys. Rev. C\/}
  {\bf 73}(5) 054309
  \urlprefix\url{http://link.aps.org/doi/10.1103/PhysRevC.73.054309}

\bibitem{THALLER1992}
Thaller B 1992 {\em The Dirac equation\/} (Berlin: Springer--Verlag)

\bibitem{JPA22:L817:1989}
Moshinsky M and Szczepaniak A 1989 {\em J. Phys. A: Math. and Gen.\/} {\bf 22}
  L817 \urlprefix\url{http://stacks.iop.org/0305-4470/22/i=17/a=002}

\bibitem{JPA40:263:2007}
Castro L~B and de~Castro A~S 2007 {\em J. Phys. A: Math. and Theor.\/} {\bf 40}
  263 \urlprefix\url{http://stacks.iop.org/1751-8121/40/i=2/a=005}

\bibitem{ABRAMOWITZ1965}
Abramowitz M and Stegun I~A 1965 {\em Handbook of Mathematical Functions\/}
  (Toronto: Dover)

\end{thebibliography}

\end{document}